\newif\ifcamready
\camreadytrue

\newif\ifhighlight
\highlighttrue

\ifcamready
\highlightfalse
\fi

\documentclass{article}
\usepackage{ijcai24}

\usepackage{times}
\usepackage{soul}
\usepackage{url}

\usepackage{natbib}

\usepackage[hidelinks]{hyperref}
\usepackage[utf8]{inputenc}
\usepackage[small]{caption}
\usepackage{graphicx}

\usepackage{siunitx}
\usepackage{amsmath}
\usepackage{amsfonts}
\usepackage{amssymb}
\usepackage{amsthm}
\usepackage{mathtools}
\usepackage{stmaryrd}
\usepackage{bm}
\usepackage{upgreek}

\usepackage{tabularx}
\usepackage{booktabs}
\usepackage{supertabular}

\usepackage{algorithm}
\usepackage{algorithmic}

\usepackage{tikz}

\usepackage{paralist}

\usepackage[switch]{lineno}

\usepackage{cleveref}
\def\T{\mathsf{T}}
\def\diag{\operatorname{diag}}

\title{Separate This, and All of these Things Around It:\\Music Source Separation via Hyperellipsoidal Queries}

\ifcamready
\author{
Karn N. Watcharasupat
\And
Alexander Lerch
\affiliations
Music Informatics Group,
Georgia Institute of Technology,
Atlanta, GA, USA
\emails
\{kwatcharasupat, alexander.lerch\}@gatech.edu
}
\else
\author{
    Anonymous Author
    \affiliations
    Anonymous Affiliation
    \emails
    anonymous@anonymous.com
}
\fi

\usepackage{microtype}

\newcommand{\ackedalex}[1]{}

\ifcamready

\else

\fi

\ifcamready
\else
\linenumbers
\fi

\ifhighlight
\newenvironment{red}{\color{red}}{}
\newenvironment{blue}{\color{blue}}{}
\else

\fi

\begin{document}

\maketitle

\begin{abstract}
Music source separation is an audio-to-audio retrieval task of extracting one or more constituent components, or composites thereof, from a musical audio mixture. 
Each of these constituent components is often referred to as a ``stem'' in literature. Historically, music source separation has been dominated by a stem-based paradigm, leading to most state-of-the-art systems being either a collection of single-stem extraction models, or a tightly coupled system with a fixed, difficult-to-modify, set of supported stems. Combined with the limited data availability, advances in music source separation have thus been mostly limited to the ``VDBO'' set of stems: \textit{vocals}, \textit{drum}, \textit{bass}, and the catch-all \textit{others}. Recent work in music source separation has begun to challenge the fixed-stem paradigm, moving towards models able to extract any musical sound as long as this target type of sound could be specified to the model as an additional query input. 
We generalize this idea to a \textit{query-by-region} source separation system, specifying the target 
based on the query regardless of how many sound sources or which sound classes are contained within it. To do so, we propose the use of hyperellipsoidal regions as queries to allow for an intuitive yet easily parametrizable approach to specifying both the target (location) as well as its spread. Evaluation of the proposed system on the MoisesDB dataset demonstrated state-of-the-art performance of the proposed system both in terms of signal-to-noise ratios and retrieval metrics. 



\end{abstract}


\section{Introduction}

Audio source separation is an audio-to-audio retrieval task with the goal of extracting one or more constituent components (``sources''), or composites thereof, from an audio mixture. Music source separation (MSS) is a subtask of audio source separation with a focus on musical audio. In MSS, the terms ``sources,'' ``stems,'' or ``instruments'' are often used interchangeably. In the broadest sense, the term ``mixture'' in audio source separation refers to the superposition of the sources or their potential transforms (e.g., via room impulse response or digital effects). In source separation literature utilizing deep learning, however, all of the transformations on the sources are usually assumed to be absorbed into the sources themselves, so that the mixture is a linear instantaneous sum of the ground truth sources.

\subsection{Fixed-Stem Music Source Separation}

Unlike the more widely studied speech source separation task, however, a musical audio mixture often contains a large number of highly correlated wide-band sources overlapping in both time and frequency, and is recorded at a high sampling rate. A common intended downstream usage is of creative nature. This means that \begin{inparaenum}[(a)]
    \item the approximate source independence assumption that underlies much of the speech source separation algorithms is almost always violated in real-world musical mixtures, and that
    \item the thresholds of acceptable quality for an MSS output are also often considerably higher than those for speech outputs. 
\end{inparaenum}

These requirements, combined with the historical peculiarities of the data availability \citep{Liutkus20172016SignalSeparation, Stoter20182018SignalSeparation}, have resulted in the MSS field being dominated by a stem-based paradigm with most work focusing on the task of extracting exactly four stems from their mixture: \textit{vocals}, \textit{drum}, \textit{bass}, and the catch-all \textit{others}, collectively known as ``VDBO.'' 
A very significant consequence of this stem-based paradigm is that most major milestone systems within MSS have either been \begin{inparaenum}[(i)]
    \item a collection of dedicated single-stem extraction models 
    \citep{%
        Uhlich2017ImprovingMusicSource, 
        Takahashi2017MultiScaleMultibandDensenets,
        Takahashi2018MMDenseLSTMEfficientCombination,
        Stoter2019OpenUnmixReferenceImplementation, 
        Kim2021KUIELabMDXNetTwoStreamNeural,
        Takahashi2021D3NetDenselyConnected,
        Luo2023MusicSourceSeparation,
        Lu2023MusicSourceSeparation}, or
    \item a tightly coupled model with a fixed set of supported stems 
    \citep{%
        Stoller2018WaveUNetMultiScaleNeural,
        Defossez2019DemucsDeepExtractor, 
        Defossez2019MusicSourceSeparation, 
        Hennequin2020SpleeterFastEfficient,
        Defossez2021HybridSpectrogramWaveform,
        Kong2021DecouplingMagnitudePhase,
        Rouard2023HybridTransformersMusic}.
\end{inparaenum} 

While recent systems, both open- and closed-source, have now begun to support some stems beyond VDBO \citep{Hennequin2020SpleeterFastEfficient, Rouard2023HybridTransformersMusic},  their implementations continue to fall into either of the above categories. Moreover, due to the data-driven nature of practically all modern systems, the exact definition of what a particular ``stem'' entails is opaque. In practice, the inclusion-exclusion criteria for a particular stem is often a \textit{post hoc} consequence of the training and dataset setups, meaning that even the developers themselves have little \textit{a priori} control over this, and potential end users even less so.

\subsection{Beyond the Fixed-Stem Paradigm}

Several attempts to support MSS beyond a fixed set of stems have been made over the years. Most of these systems rely on some form of query-by-example or conditioning. As early as 2019, although still with VDBO outputs, \citet{Meseguer-Brocal2019ConditionedUNetIntroducingControl} used a U-Net with a single decoder to perform MSS, conditioned via multiple feature-wise linear modulation (FiLM) modules \citep{Perez2017FiLMVisualReasoning}.
This approach and its variations are common with conditional MSS works within and beyond VDBO, as well as in some cross-domain source separation works \citep{
Lee2019AudioQuerybasedMusic,
Li2019CreatingMultitrackClassical,
Choi2021LaSAFTLatentSource,
Jeong2021LightSAFTLightweightLatent,
Lin2021UnifiedModelZeroshot,
Gfeller2021OneShotConditionalAudio,
Slizovskaia2021ConditionedSourceSeparation,
Chen2022ZeroshotAudioSource,
Liu2022SeparateWhatYou, 
Liu2023SeparateAnythingYou, Kong2023UniversalSourceSeparationa}.  


Taking the idea of conditional MSS further, some recent works have begun to more substantially challenge the stem-based paradigm. These works are moving beyond a fixed-stem or stem-based paradigm where a single model can extract any source based on some form of additional input that is no longer limited to stem labels. 
\citet{Manilow2022SourceSeparationSteering} used gradient ascent to steer a pre-trained MSS model to extract target sources based on auto-tagging predictions. \citet{Petermann2023HyperbolicAudioSource} used a very low-dimensional hyperbolic embedding space to allow similarity-based source separation based on the selected subset of the Poincaré disk. Although this work had a very intuitive interface, the use of very low-dimensional hyperbolic space  severely limited the separation fidelity. 
\citet{Wang2022FewShotMusicalSource} used a single FiLM conditioning in the U-Net bottleneck to perform few-shot MSS, although its performance was weaker than contemporary models. A similar approach was later taken in Banquet \citep{Watcharasupat2024StemAgnosticSingleDecoderSystem}, a query-based approach  achieving competitive performance to the state-of-the-art fixed-stem HTDemucs \citep{Rouard2023HybridTransformersMusic}. 
Similar to the system proposed by \citet{Wang2022FewShotMusicalSource}, however, a major limitation of Banquet is that the querying method only specifies a single \textit{point} in the embedding space, around which the relevant signals should be extracted, with no control over the ``broadness,'' spread, or specificity of the query. 


\subsection{Main Contributions}

In this work, we move further beyond the stem-based paradigm, meaning that the extraction target is no longer necessarily treated in a stem-based manner. Instead, we propose to develop a system that can extract any arbitrary target, composite or single-source, as specified by the query. To do so, we generalize query-based MSS beyond \textit{point} query to \textit{region} query, that is, to provide the model with both the ``location'' and the ``spread'' of the query. This is loosely similar to the idea of bounding-box queries in image segmentation \citep{Kirillov2023SegmentAnything}, but in the abstract embedding space.

We propose using hyperellipsoidal regions as queries, which have a close relationship to the highly generalizable multivariate Gaussian distribution. Specifically, we extend the Banquet model  \citep{Watcharasupat2024StemAgnosticSingleDecoderSystem} to support a query-by-region operation, resulting in a more customizable source separation method that also supports zero- to few-shots use cases.
To the best of our knowledge, our work is the only other system to perform query-based source separation with region query apart from the aforementioned hyperbolic system   \citep{Petermann2023HyperbolicAudioSource}. 

In addition, given the difficulty of evaluating the retrieval performance of query-based MSS systems with a large number of stems and corresponding large timbral diversity, we also introduce a method to evaluate our audio-to-audio MSS system as a retrieval system. This is done by using simple least-squares projection to approximate the presence or absence of a particular source in the model output (i.e., the collection of retrieved items). Given a known target set of sources, we can then determine the extent to which the model has retrieved target sources (i.e., relevant items) or interferences (i.e., irrelevant items) from the mixture (i.e., a collection of all items), despite the system output being that of a superposition of all retrieved signals.


\section{Proposed System}



\begin{figure}
    \centering
    \includegraphics[width=\columnwidth]{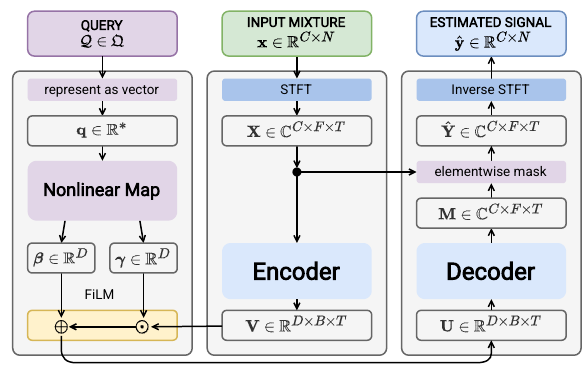}
    \caption{Overview of the Proposed System}
    \label{fig:overview}
\end{figure}

The overall system in this work is that of a complex-valued time-frequency (TF) masking source separation, with a single query-based embedding adaptation at the bottleneck.\ackedalex{That does sentence doesn't strike me as defining an architecture at all.} The benefit of using a TF-domain masking technique for source separation is that there is a direct connection to the time-domain time-varying convolutive filters, and that the model is incapable of ``hallucination'' --- any information outputted by the model must have already existed in the input. The overall system is visualized in \Cref{fig:overview}.

\ackedalex{I am missing a simple flow-chart here, showing encoder, single decoder, and query with projection in some way. It might seem trivial, but I still think it's necessary.}

The general operation of any such system can be described as follows. Let $\mathbf{x} \in \mathbb{R}^{C \times N}$ be the input audio mixture with $C$ channels. Let $\mathbf{X} = \operatorname{STFT}(\mathbf{x}) \in \mathbb{C}^{C \times F \times T}$, where $\operatorname{STFT}$ is the short-time Fourier transform operator, $F$ is the number of non-redundant frequency bins, and $T$ is the number of temporal frames. Let $\mathcal{Q} \in \mathfrak{Q}$ be a query representation in some space $\mathfrak{Q}$. The overall system can be described by the sequence of operations:
\begin{align}
\begin{aligned}
    \mathbf{X} &= \operatorname{STFT}(\mathbf{x}) && \in\mathbb{R}^{C \times F \times T}\\
    \mathbf{V} &= \operatorname{Enc}(\mathbf{X}) && \in\mathbb{R}^{D \times B \times T}\\
    \mathbf{U} &= \operatorname{Cond}(\mathbf{V}, \mathcal{Q}) && \in\mathbb{R}^{D \times B \times T}\\
    \mathbf{M} &= \operatorname{Dec}(\mathbf{U}) && \in\mathbb{C}^{C \times F \times T}\\
    \hat{\mathbf{Y}} &= \mathbf{M} \circ \mathbf{X} && \in\mathbb{C}^{C \times F \times T}\\
    \hat{\mathbf{y}} &= \operatorname{iSTFT}(\hat{\mathbf{Y}}) && \in\mathbb{R}^{C \times N}
\end{aligned}
\end{align}
with the mixture encoder $\operatorname{Enc} \colon \mathbb{C}^{C \times F \times T} \mapsto \mathbb{R}^{D \times B \times T}$, the mask decoder $\operatorname{Dec} \colon \mathbb{R}^{D \times B \times T} \mapsto \mathbb{C}^{C \times F \times T}$, and a query conditioning module $\operatorname{Cond} \colon (\mathbb{R}^{D \times B \times T}, \mathfrak{Q}) \mapsto \mathbb{R}^{D \times B \times T}$. We will refer to $\mathbf{V}$ as the mixture embedding, $\mathbf{U}$ as the conditioned embedding, $\mathbf{M}$ as the mask, and $\hat{\mathbf{Y}} \leftrightarrow\hat{\mathbf{y}}$ as the estimated output. 

\subsubsection{The Query and the Query Space} 
The query space $\mathfrak{Q}$ can be defined in different ways. On the one hand,  $\mathfrak{Q}$ can simply be a space of one-hot vectors directly representing class labels \citep{Watcharasupat2024FacingMusicTackling}. On the other hand, it can be extracted from a short audio signal by a pretrained feature extractor such as PaSST \citep{Koutini2022EfficientTrainingAudio} to obtain the query embedding \citep{Watcharasupat2024StemAgnosticSingleDecoderSystem}.
While the former model learned each class directly from the training set, the latter is in principle open to `unseen' queries but entirely dependent on a single short audio example, or the centroid of multiple examples. Neither method allows for the user control of variable levels of ``broadness'' or specificity.

\ackedalex{I thought this was a method description? It reads like related work?! Seems like wasting a lot of space on stuff that's already published.}

\subsubsection{FiLM Conditioning}
Inspired by other query-based source separation systems \citep{Watcharasupat2024FacingMusicTackling,Watcharasupat2024StemAgnosticSingleDecoderSystem}, the conditioning is performed via FiLM at the bottleneck. By representing the query $\mathcal{Q}$ as a vector, a small fully-connected network (FCN) can be used to map the vector representation to FiLM parameters $\bm{\gamma}, \bm{\beta} \in \mathbb{R}^{D}$. 
These vectors are then used to condition the mixture embedding, yielding
\begin{align}
    \mathbf{U} &= \bm{\gamma} \circ \mathbf{V} + \bm{\beta}
\end{align}
as the resultant embedding for downstream decoding. 

\subsubsection{Normalization and Reparametrization}
In this work, the mixture encoder and the mask decoder largely follow those of \citet{Watcharasupat2023GeneralizedBandsplitNeural}, but with a few additional optimizations to reduce instability during training. Specifically, we introduced
\begin{inparaenum}[(i)]
\item global normalization and denormalization similar to that of Demucs \citep{Defossez2019DemucsDeepExtractor, Defossez2021HybridSpectrogramWaveform}, to ensure that the rest of the model is trained independently of the input signal levels,
and
\item reparametrization of the linear layers in the bandsplit and mask estimation modules using the weight normalization \citep{Salimans2016WeightNormalizationSimple}, which has been shown to improve model convergence \citep{Wu2020ImplicitRegularizationConvergence}.
\end{inparaenum} 

\subsection{Hyperellipsoid as Query}\label{sec:hyper}

One method to provide the end-user (e.g., musicians) with some level of control over the  ``broadness'' of their query is to allow direct specification of the subset of the abstract space to extract. \citet{Petermann2023HyperbolicAudioSource} have proposed a low-dimensional hyperbolic representation to enable this. However, their method requires the entire system to also follow this low-dimensional modeling, thus limiting the achievable fidelity. Crucially, however, the method proposed by \citet{Petermann2023HyperbolicAudioSource} allows the user to draw an ``ellipse,'' or a union of multiple ``ellipses,'' over the Poincaré disk representation of the embedding space.\footnote{Demo: \href{https://youtu.be/RKsAMb9z70Y}{youtu.be/RKsAMb9z70Y}. Last accessed: 21 Jan 2025.}

Inspired by this intuitive control method, we propose the use of a hyperellipsoid in the query embedding space as a query to the model. That is, the model is tasked to obtain all sound components within the mixture whose embeddings lie within the region enclosed by the query hyperellipsoid. 

\subsubsection{Query by Region}

The concept of source separation using query by region can be formalized as follows. Let $\mathfrak{S} \coloneqq \{\mathbf{s}_i \in \mathbb{R}^{C \times N}\}$ be the set of sound sources contained within the mixture $\mathbf{x}$. As introduced earlier, $\mathbf{x}$ is the input mixture, which can now be more formally defined as
\begin{align}
    \mathbf{x} &\coloneqq \sum_{\mathbf{s}_i \in \mathfrak{S}} \mathbf{s}_i.
    \label{eq:mix}
\end{align}
Let $\operatorname{Emb} \colon \mathbb{R}^{C \times N} \mapsto \mathbb{R}^P$ be some embedding operation (e.g. PaSST) on the sound source signals, and let $\mathcal{Q}$ be some closed region in $\mathbb{R}^P$, acting as the query. The target signal  $\mathbf{y}$ of the extraction with input $\mathbf{x}$ and query $\mathcal{Q}$ is thus defined as the sum of all the sources whose corresponding embeddings lie within $\mathcal{Q}$,
\begin{align}
    \mathbf{y} &= \sum_{\mathbf{s}_i \in \mathfrak{Y}} \mathbf{s}_i,
    \label{eq:target}
\end{align}
where $\mathfrak{Y} \coloneqq \{ \mathbf{s}_i \in \mathfrak{S} \colon \operatorname{Emb}(\mathbf{s}_i) \in \mathcal{Q}\}$. As mentioned earlier, this work will focus on the case where $\mathcal{Q}$ is a region enclosed by a hyperellipsoid in the embedding space.

\subsubsection{Region Bounded by a Hyperellipsoid}
\ackedalex{Should we add a quick visualization here? This is the part that we should use the paper space for.}

\begin{figure}[t]
    \centering
    \includegraphics[width=0.75\columnwidth]{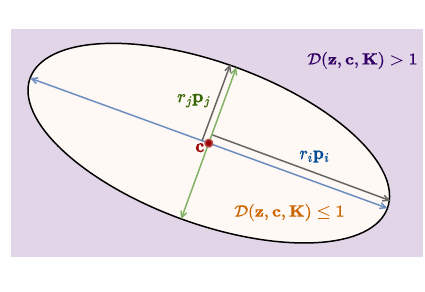}
    \caption{A two-dimensional cross-section of a hyperellipsoid in $\mathbb{R}^K$, $K\ge 3$, via a hyperplane $\{\mathbf{w} \in \mathbb{R}^{K} \colon \mathbf{p}_k^\T(\mathbf{w}-\mathbf{c}) = 0,\ \forall k\ne i,j \}$.}
    \label{fig:ellipse}
\end{figure}

Generally, the region bounded by a hyperellipsoid in $\mathbb{R}^P$ can be described by
\begin{equation}
    \mathcal{Q}(\mathbf{c}, \mathbf{K}) \coloneqq \left\{
        \mathbf{z} \in \mathbb{R}^P \colon \mathcal{D}(\mathbf{z}, \mathbf{c}, \mathbf{K}) \le 1
    \right\}
\end{equation}
where $\mathbf{c} \in \mathbb{R}^{P}$ is the center of the hyperellipsoid,  $\mathbf{K} \eqqcolon \mathbf{P}\mathbf{\Lambda}\mathbf{P}^\T$ is a positive-definite matrix in $\mathbb{R}^{P \times P}$, $\mathbf{P}$ is an orthonormal matrix in $\mathbb{R}^{P \times P}$ defining the principal axes of the hyperellipsoid, and $\mathbf{r} \in \mathbb{R}^{P}_{+}$ is the semi-axis lengths, such that $\mathbf{\Lambda} = \diag(\mathbf{r})^2$, and
\begin{equation}
    \mathcal{D}(\mathbf{z}, \mathbf{c}, \mathbf{K})
    = (\mathbf{z} - \mathbf{c})^\T \mathbf{K}^{-1} (\mathbf{z} - \mathbf{c}) \in \mathbb{R}_{0}^+
\end{equation}
is the Mahalanobis distance. In practice, $\mathbf{K}^{-1}$ above might not be well-conditioned, and is thus replaced by $\mathbf{K}^{\dagger} = \mathbf{P}\mathbf{\Lambda}^\dagger\mathbf{P}^\T$, where
\begin{align}
    (\mathbf{\Lambda}^\dagger)_{i,i} &= r_i^{-2} \ \text{if}\ r_i \ge \epsilon,\ \text{else}\ 0,
\end{align}
for some small $\epsilon$. A two-dimensional cross-section of a hyperellipsoid in $\mathbb{R}^K$, $K\ge 3$, via a hyperplane $\{\mathbf{w} \in \mathbb{R}^{K} \colon \mathbf{p}_k^\T(\mathbf{w}-\mathbf{c}) = 0,\ \forall k\ne i,j \}$. is visualized in \Cref{fig:ellipse}.

The choice to use a hyperellipsoid as the geometric query is due to its close relation to its probabilistic ``counterpart,'' the multivariate Gaussian distribution, which can approximate many unimodal distributions in high-dimensional space. Additionally, the hyperellipsoidal query lends itself naturally to multimodal extension (i.e., Gaussian mixtures) in a more complex querying scenario beyond the scope of the present work.

\subsubsection{Dimensionality Reduction}
Using the PaSST embedding space as the query space is a natural choice as the PaSST space has been shown to be highly discriminative for tasks such as musical instrument classification \citep{Koutini2022EfficientTrainingAudio}.
The raw 768-dimensional space, however, is not only significantly higher-dimensional than the bottleneck space with $D=128$, but also not empirically full-rank. This leads to unnecessary complexity, memory consumption, and~---most importantly---~severe numerical stability issues. To alleviate this, we perform principal component analysis (PCA) on the PaSST embedding space to reduce it to a $D$-dimensional space. The explained variance of this 128-dimensional reduced space is \SI{91.8}{\percent} \ackedalex{Does it makes sense to give a number here without specifying $D$?} based on the training set. From here, we can assume $P = D$. 

\subsubsection{Vector Representation of the Query}

In this work, the hyperellipsoid $\mathcal{Q}(\mathbf{c}, \mathbf{K})$ is represented somewhat naively as an $D(D+3)/2$-dimensional vector $\mathbf{q}$, by exploiting the symmetry of $\mathbf{K}$,
\begin{align}
    \mathbf{q}^\T &= \begin{bmatrix}\mathbf{c}^\T & \operatorname{tril}(\mathbf{K})^\T\end{bmatrix},
\end{align}
where $\operatorname{tril}$ returns lower triangular elements of its input matrix as a vector. The mapping from $\mathbf{q}$ to the FiLM parameters is done using a small FCN as in MSS Banquet. 

\subsection{Loss Function}

\subsubsection{Reconstruction Loss}

The reconstruction loss used in this work is the multi-domain multichannel L1SNR loss following \citet{Watcharasupat2024StemAgnosticSingleDecoderSystem}. 

\ackedalex{You are just repeating previous work. Why is this relevant here?}

\subsubsection{Level-Matching Regularization}
In order to alleviate the issues with near-silent output raised in \citet{Watcharasupat2024StemAgnosticSingleDecoderSystem}, we introduce a simple level-matching regularization $\mathcal{R}(\hat{\mathbf{y}}; \mathbf{y})$. Define
$L \coloneqq \operatorname{dBRMS}(\mathbf{y})$ and $\hat{L} \coloneqq\operatorname{dBRMS}(\hat{\mathbf{y}})$. We have
\begin{align}
    \mathcal{R}(\hat{\mathbf{y}}; \mathbf{y}) &= \lvert \hat{L} - L\rvert, 
\end{align}
with an adaptive weighting
\begin{align}
    \lambda(\hat{\mathbf{y}}; \mathbf{y})
    &= \lambda_{0} + \eta \cdot \Delta\lambda \cdot \underset{[0,1]}{\operatorname{clamp}}\left[\dfrac{\mathcal{R}(\hat{\mathbf{y}}; \mathbf{y})}{L - L_{\min}}\right],
\end{align}
where $\eta \coloneqq \mathbb{I}\llbracket L > \max(\hat{L}, L_{\min})\rrbracket \in \{0, 1\}$, $L_{\min}$ is the minimum RMS level in dB, $\lambda_0$ is the minimum weight, $\Delta \lambda$ is the weight range, giving $\lambda_0 + \Delta \lambda$ as the maximum weight. The principle behind the adaptive weighting is to increasingly penalize under-level predictions, while only minimally penalizing over-level predictions.

\subsubsection{Overall Loss}

The complete loss function is thus
\begin{align}
    \mathcal{J}(\hat{\mathbf{s}}; \mathbf{s}) &= \mathcal{L}(\hat{\mathbf{s}}; \mathbf{s}) + \operatorname{sg}[\lambda(\hat{\mathbf{s}}; \mathbf{s})] \cdot \mathcal{R}(\hat{\mathbf{s}}; \mathbf{s}),
\end{align}
where $\operatorname{sg}[\cdot]$ is the stop-gradient operator. Our repeated preliminary experiments have indicated that the proposed model eventually collapses during training without the level-matching regularization.

\section{Experimental Setup}

\subsection{Data}

This work utilizes MoisesDB \citep{Pereira2023MoisesDBDatasetSource}, a multi-stem music source separation dataset. For comparability, the data splits follow that of \citet{Watcharasupat2024StemAgnosticSingleDecoderSystem}. Abbreviations of the stem names from MoisesDB are shown in \Cref{tab:instrument_abbreviations} of the Appendix.

In theory, one desirable approach for training the proposed system would be to expose it to a maximum diversity of queries, in terms of included target sounds, centers, and semi-axis lengths. In practice, however, we continue to be limited by the use of supervised learning, as is the case for nearly all high-fidelity source separation systems to date. This means that the granularity of the included target sounds is dictated by what the dataset provides. This limitation, combined with the sparseness of high-dimensional space, means that naive on-the-fly random sampling of queries is extremely inefficient. The heuristic approach to form a set of valid queries is outlined below.
    
\subsubsection{Precomputing Valid Queries}

\begin{figure}
    \centering
    \includegraphics[width=\columnwidth]{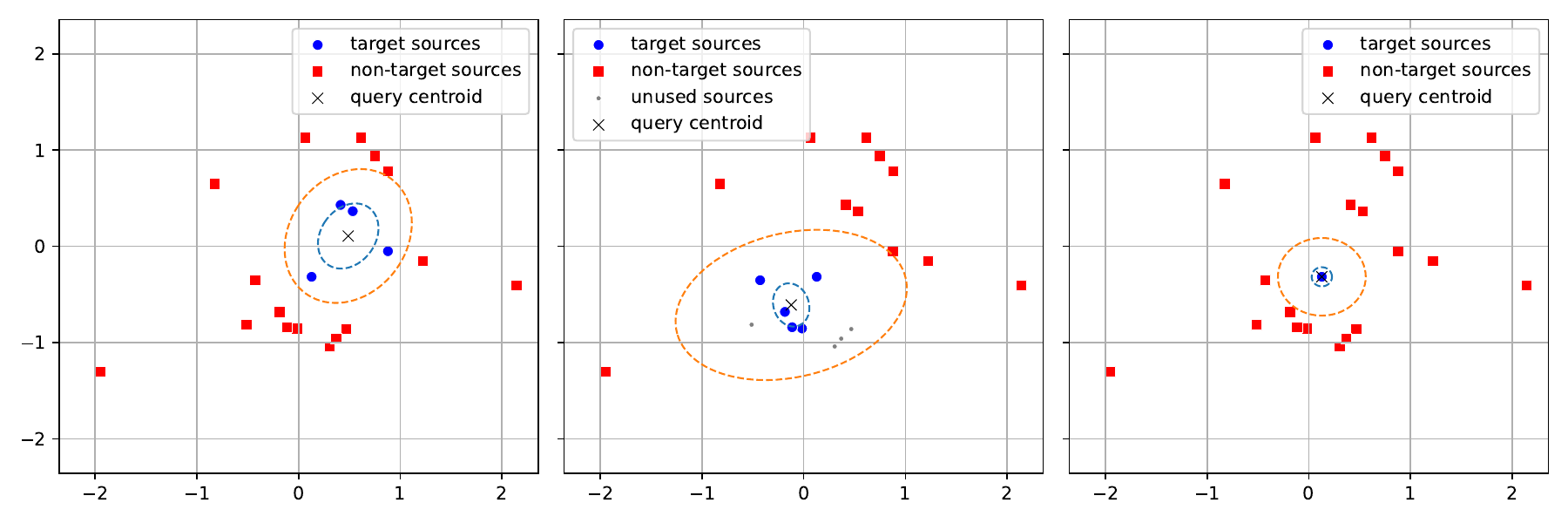}
    \caption{Simplified two-dimensional representation of some possible training queries from the same training data point, meaning that all source embeddings are the same across the three cases. For each case, the valid query with the same input mixture and the target is any ellipse interpolating between the inner (blue) and outer (red) dotted lines.   
    \textbf{Left:} The target is a composite of four sources. The input mixture is a composite of all available sources.
    \textbf{Center:} The target is a composite of five sources. Not all available sources are used in the mixture.
    \textbf{Right:} The target is a single source. The input mixture is a composite of all available sources.} 
    \label{fig:enter-label}
\end{figure}

For each 10-second sliding-window chunk within a track (``clip''), we iterate through all possible subsets of available sources. For each subset acting as the set of target sources, we find the smallest enclosing hyperellipsoid that contains all of their embeddings. We eliminate any non-target source with its embedding lying within this hyperellipsoid from the mixture. We then find the largest excluding hyperellipsoid with the same center, such that the embeddings of all the remaining target and non-target sources all lie outside it. Any hyperellipsoid formed by interpolating between the enclosing and the excluding hyperellipsoid would thus specify the same target given the corresponding input mixture. More precise mathematical details are described below.

Let $\mathfrak{U}$ be the set of available sources within each clip, as provided by the dataset. To avoid silent target sources, within each clip, only sources that are at least \SI{-48}{dBRMS} in level were considered to be available. Compute the set of corresponding embeddings $\mathfrak{Z} \eqqcolon \{\mathbf{z}_j\}$.
For every possible proper subset $\pi \subset \mathfrak{U}$ corresponding to $\mathfrak{Z} \subset \mathfrak{Z}$, we estimate the mean $\mathbf{c}$ and covariance $\mathbf{\Sigma} = \mathbf{P}_{i}\mathbf{\Lambda}\mathbf{P}^{\T}_{i}$ of $\mathfrak{Z}$, where $\mathbf{P}$ is orthorgonal and $\mathbf{\Lambda} = \diag(\bm{\uplambda})$. If $|\mathfrak{Z}| = 1$, $\mathbf{K}$ is theoretically zero. In practice, we set $\mathbf{K} = \delta\mathbf{I}$ with a small $\delta$. Otherwise, $\mathbf{K}$ set $\mathbf{K} \coloneqq \kappa \mathbf{\Sigma}$, where
\begin{align}
    \kappa \coloneqq \textstyle\max_{\mathbf{z} \in \mathfrak{Z}}\ 
    \mathcal{D}(\mathbf{z}, \mathbf{c}, \mathbf{\Sigma}),
\end{align}
so that $\mathbf{z}_j \in \mathfrak{E} \coloneqq \mathcal{Q}(\mathbf{c}, \mathbf{K})$, $\forall \mathbf{z}_j \in \mathfrak{Z}$. Define the non-target embedding set as $\mathfrak{Z}' \coloneqq {\mathfrak{Z}\setminus(\mathfrak{Z} \cup \mathfrak{E})}$. Estimate the covariance $\mathbf{\Sigma}'$ of $\mathfrak{Z}'$ about $\mathbf{c}$. Set $\mathbf{K}' \coloneqq \kappa' \mathbf{\Sigma}'$, where
\begin{align}
    \kappa' \coloneqq \textstyle\min_{\mathbf{z} \in \mathfrak{Z}'}\ 
    \mathcal{D}(\mathbf{z}, \mathbf{c}, \mathbf{\Sigma}'),
\end{align}
Assuming that non-zero degenerate eigenvalues of $\mathbf{\Sigma}'$ are extremely rare in practice, we can use the simultaneous orthogonal diagonalization property of positive definite matrices to compute
\begin{align}
    \bm{\uplambda}^{\perp} \coloneqq \max\{\diag(\mathbf{P}^{\T}_{i} \mathbf{K}' \mathbf{P}_{i}),\, \bm{\uplambda}\}. \label{eq:simuldiag}
\end{align}
Set $\mathfrak{Z}^\perp \coloneqq \mathfrak{Z}' \setminus \mathcal{Q}(\mathbf{c}, \mathbf{P}_{i}\mathbf{\Lambda}^\perp\mathbf{P}^{\T})$ to catch any edge cases from numerical issues. Store the centroid $\mathbf{c}$, the inclusion radii $\mathbf{r} \coloneqq \bm{\uplambda}^{\circ {1/2}}$, the exclusion radii $\mathbf{r}^\perp \coloneqq (\bm{\uplambda}^{\perp})^{\circ 1/2}$, the principal axes $\mathbf{P}$, the target source indices $\pi$, and the non-target source indices $\pi^\perp$.

During training, draw $\tilde{\mathbf{r}} \sim \operatorname{Unif}(\mathbf{r}, \mathbf{r}^\perp)$. The query is then defined by $\mathcal{Q}(\mathbf{c}, \mathbf{P} \diag(\tilde{\mathbf{r}})^2 \mathbf{P}^{\T})$.
All sources in $\pi \cup \pi^{\perp}$ are augmented. The target $\mathbf{y}$ is formed by linearly summing augmented sources from $\pi$, while the input mixture $\mathbf{x}$ is formed by linearly summing the \textit{same} set of augmented sources from $\pi \cup \pi^{\perp}$.

For the validation set, the radii are formed deterministically with $\tilde{\mathbf{r}} \coloneqq (\mathbf{r}+ \mathbf{r}^\perp)/2$. No augmentation was performed to the validation set.

\subsection{Evaluation Metrics}

MSS systems are audio-to-audio retrieval systems. This means that its evaluation has to involve both audio quality evaluation and retrieval quality evaluation. That is, we ask, \begin{inparaenum}[(1)]
    \item ``Did the model output the correct sources'', and
    \item ``Are the model output signals of a high quality?''. 
\end{inparaenum}
Given that MSS systems output audio \textit{signals}, not logits, confidences, or ranks, these two questions are not easily disentangled in practice.

One approach, that was extremely popular for tackling these evaluations, is that of the BSSEval toolkit \citep{Vincent2007FirstStereoAudio}. At the time of its development, however, source separation research was largely focused on array source separation of speech signals in a \textit{multichannel linear convolutive} mixture, usually with a maximum of four sources \citep{Vincent2012SignalSeparationEvaluation}. Unsurprisingly, the {de facto} standard technique to decompose the error signal into distortion, interference, and artifact signals was designed in that historical context. Despite the enormous usefulness of this decomposition and the resulting metrics, its computation involves orthogonal subspace projections onto the span of the shifted source signals. Amongst other issues \citep{LeRoux2019SDRHalfbakedWell,Scheibler2022SDRMediumRare}, this projection is numerically unstable and unreliable when there are a large number of highly correlated sources. This limitation has already caused some problems in a four-stem VDBO MSS setup \citep{Stoter20182018SignalSeparation}, where each output is usually predesignated to a particular stem. In a query-based system like ours, these problems are amplified, rendering the BSSEval metrics, if numerically computable at all, uninformative for evaluating our system.  

However, we still want to answer the two evaluation questions above. To do so, we first rely on the signal-to-noise ratio, a standard audio metric. We then developed a method to approximate the extent to which a source has been retrieved by the model, allowing the system to be evaluated using more convention retrieval metrics.


\subsubsection{Signal-to-Noise Ratio}

In alignment with recent MSS works and the Music Demixing Challenges \citep{Mitsufuji2022MusicDemixingChallenge, Fabbro2024SoundDemixingChallenge}, we evaluate our system using the signal-to-noise ratio (SNR), which is defined by
\begin{align}
    \operatorname{SNR}(\hat{\mathbf{y}}; \mathbf{y}) = 10 \log_{10} \dfrac{\| \mathbf{y}\|_F^2 + \xi}{\|\hat{\mathbf{y}} - \mathbf{y}\|_F^2+ \xi},
\end{align} 
where $\hat{\mathbf{y}}$ is the test signal and $\mathbf{y}$ is the reference signal. We use $\xi = \num{e-6}$ for stability. 

SNR is a relatively unforgiving metric in that any sample-level deviation of $\hat{\mathbf{y}}$ from $\mathbf{y}$ is treated as an error, regardless of whether this is due to distortions on the (correct) target source,  interferences from non-target sources, or just spatial errors. In other words, retrieval errors \textit{will} decrease SNR, but retrieval errors cannot be detected from a decrease in SNR without further inspection.
Note also that the consequence of this SNR definition is that a silent $\hat{\bm{y}}$ will produce zero SNR. It is thus important to consider the output levels of the predictions together with the SNR, especially when the SNR is close to zero. 

\begin{figure*}[t]
    \centering
    \includegraphics[width=\linewidth]{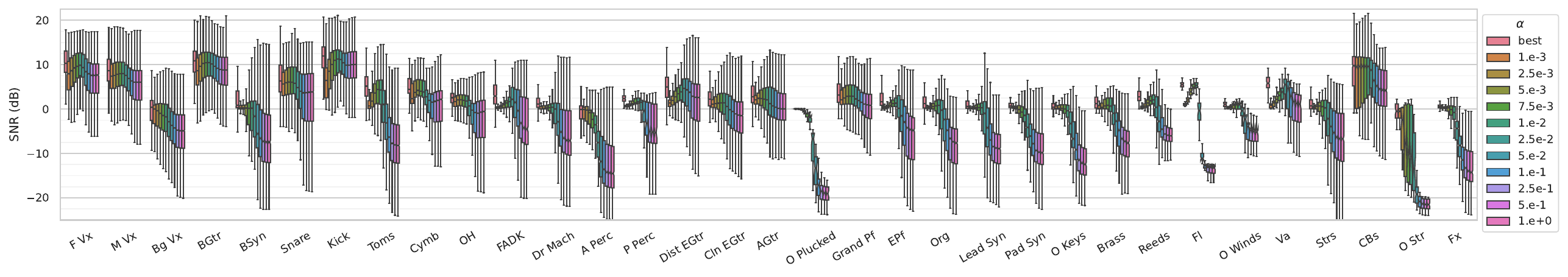}
    \caption{Single-source queries: SNR (dB) distribution by target ``stem'' over query scale factors, $\alpha$, and the clip-wise best factor. }
    \label{fig:1s-snr}
\end{figure*}

\subsubsection{Retrieval Metrics}

Given the query-based nature of our system, it is important to still evaluate the retrieval performance of the system in a manner that is as independent from the signal quality as possible. Using the same definitions as in \Cref{eq:mix,eq:target}, we additionally define the set of non-target sources within each clip as $\mathfrak{Y}^{\perp} \coloneqq \mathfrak{S} \setminus \mathfrak{Y}$. Now, given the corresponding estimated signal $\hat{\mathbf{y}}$, we solve for
\begin{align}
    \min_{%
        \tilde{\mathbf{\upphi}} 
            \in \mathbb{R}^{|\mathfrak{Y}|},
        \tilde{\mathbf{\upphi}}^{\perp}
            \in \mathbb{R}^{|\mathfrak{Y}^{\perp}|}
    } \|\hat{\mathbf{y}} - \tilde{\mathbf{y}}(\tilde{\mathbf{\upphi}}, \tilde{\mathbf{\upphi}}^\perp)\|_F^2
\end{align}
where 
\begin{align}
    \tilde{\mathbf{y}}(\tilde{\mathbf{\upphi}}, \tilde{\mathbf{\upphi}}^\perp) &= \sum_{\mathbf{s}_i \in \mathfrak{Y}}\tilde{\phi}_i\mathbf{s}_i + \sum_{\mathbf{s}_j^{\perp} \in \mathfrak{Y}^{\perp}}\tilde{\phi}_j^{\perp}\mathbf{s}_j^{\perp}.
\end{align}
The weight vectors $\tilde{\mathbf{\upphi}} \in \mathbb{R}^{|\mathfrak{Y}|}$ and $\tilde{\mathbf{\upphi}}^{\perp}\in \mathbb{R}^{|\mathfrak{Y}^{\perp}|}$ can be obtained using any standard least-squares solver. From here, we normalize elementwise
\begin{align}
    \hat{\mathbf{\upphi}} &\coloneqq \min\{\bm{1}, |\tilde{\mathbf{\upphi}}|_{\circ}\}, \qquad\hat{\mathbf{\upphi}}^{\perp} \coloneqq \min\{\bm{1}, |\tilde{\mathbf{\upphi}}^{\perp}|_{\circ}\}.
\end{align}
The vectors $\hat{\mathbf{\upphi}}$ and $\hat{\mathbf{\upphi}}^{\perp}$ can then be treated as classification ``scores'' with the corresponding ``true'' values $\mathbf{\upphi} = \mathbf{1}$ and $\mathbf{\upphi}^\perp = \mathbf{0}$. Standard retrieval metrics can now be used to evaluate our system. In this work, we will use accuracy, precision, recall, F1, average precision (AP), mean average precision (mAP), and are under the receiver operating characteristics curve (ROCAUC). 

\section{Results}

To evaluate the performance of the system, we chunked each track
within the test set into clips using a sliding window of length \SI{10}{\second} and a stride of \SI{1}{\second}. The system was then evaluated using all possible queries, precomputed in the same way as the training and validation sets. For comparability, the model outputs are evaluated in a stem-based manner. It should be emphasized again, however, that the proposed system was never provided the classes of the stems to be extracted during the training or testing.

\begin{figure}[t]
    \centering
    \includegraphics[width=\columnwidth]{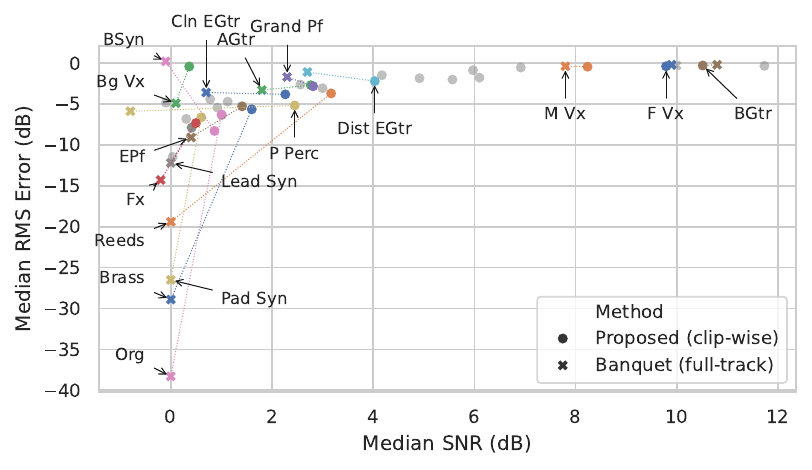}
    \caption{Single-source queries: Plot of the median SNRs (dB) and against median RMS errors (dB) of the proposed method and Banquet (\textsc{q:all}, TE+DA variant). Note again that our method was evaluated clip-wise while Banquet was evaluated over the full track.}
    \label{fig:snr-rms}
\end{figure}

\subsection{Single-Source Queries}

Given that the inclusion radii, $\mathbf{r}$, of a single-source query is theoretically the $\bm{0}^+$ vector, finding the optimal query for the single-source cases can be more challenging than the multi-source cases. For each clip, we query the system using hyperellipsoids centered at the centroid of the target embedding and with radii $\tilde{\mathbf{r}} = \alpha\mathbf{r}^\perp$, for various $\alpha \in [\num{e-3}, 1]$. The distribution of the SNR over each target ``stem'' with various $\alpha$'s are shown in \Cref{fig:1s-snr}. Using the clip-wise best $\alpha$ and considering only the clips where all sources are included in the mixture, the median SNRs and root-mean-square (RMS) errors of the proposed method are shown in \Cref{fig:snr-rms},\footnote{See \Cref{tab:snr} for the tabular version of \Cref{fig:snr-rms}.} and compared against the reported results in Banquet (\textsc{q:all}, TE+DA variant) from \citet{Watcharasupat2024StemAgnosticSingleDecoderSystem}, where applicable. 
Note, however, that due to the differences in the evaluation setups, the SNR reported by \citet{Watcharasupat2024StemAgnosticSingleDecoderSystem} was computed over the full track using overlap-add, while ours is computed clip-wise, thus the comparison should only be taken as a rough gauge of relative performance.

Overall, our method performed on par with Banquet on \textit{lead vocals} and \textit{bass guitar}, while performing better on the remainder of the source classes. The most significant improvement over Banquet lies in the performance of sources belonging to ``long-tail'' instrument classes (\textit{organs}, \textit{synths}, \textit{brass}, \textit{reeds}, \textit{strings}). As noted by \citet{Watcharasupat2024StemAgnosticSingleDecoderSystem}, Banquet suffered from collapsed outputs on these sources, as evidenced by the severely negative RMS errors and zero SNRs. While our method still has median RMS errors hovering around \SI{-6}{dB}, this, combined with the positive SNR, indicated that the proposed method can now extract these sources better than Banquet. Interestingly, \textit{flute}, viola, and \textit{contrabass} performed better than expected. \textit{Flute} only occurs as a target in \SI{0.4}{\percent} of the training samples, \textit{contrabass} in \SI{1.0}{\percent}, and \textit{viola} is not present in the training set at all. Stem classes similar to \textit{viola} and \textit{contrabass} in the training set are also relatively rare, with \textit{cello} at \SI{1.6}{\percent} and cello section at \SI{0.1}{\percent}. 
\ackedalex{Should I be able to find these in Figure 5? Because I can't. Or have you moved to a different plot in the discussion? I think it might be better to structure the discussion figure by figure, that would make it easier to follow. Otherwise the reader has to guess where to look at to see what you are talking about.}

\begin{figure}[t]
    \centering
    \includegraphics[width=\columnwidth]{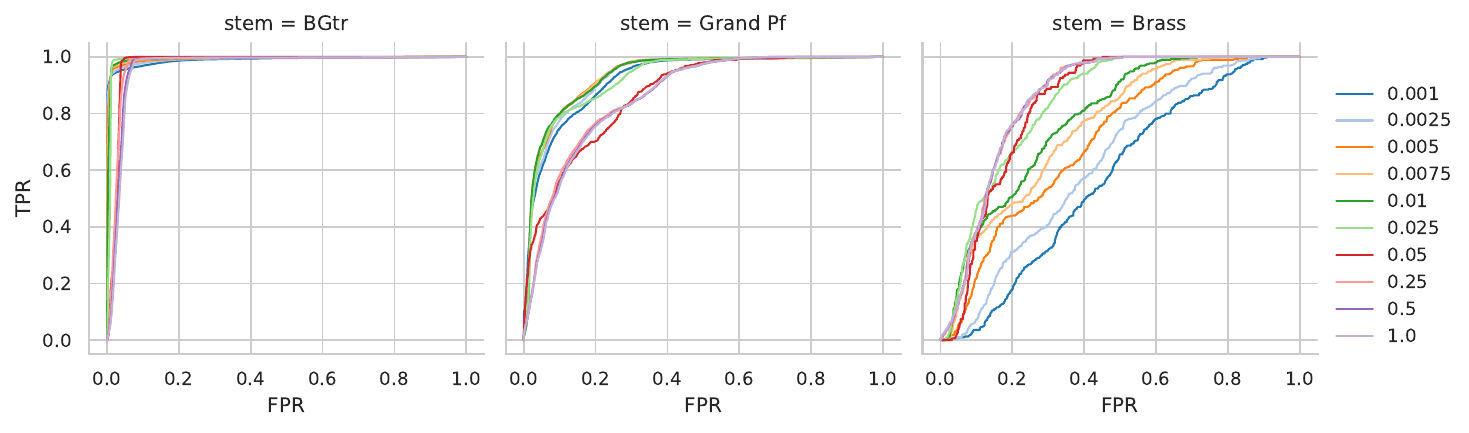}
    \caption{Single-source queries: ROC for lead bass guitar, grand piano, and brass, over different $\alpha$.}
    \label{fig:roc}
\end{figure}

In terms of $\alpha$, we observe that the query width can have significant effects on the model performance. Example effects of $\alpha$ can be seen in the ROC curves in \Cref{fig:roc}. See \Cref{fig:all-roc} for one with every source class.
In the example on the left (\textit{bass guitar}), the model is relatively insensitive to $\alpha$. For grand piano (middle), $\alpha \le 0.025$ (outer cluster of lines) performed similarly amongst one another, but increasing $\alpha \ge 0.05$ (inner cluster) resulted in a marked drop in ROCAUC. On the other hand, brass (right) performed similarly amongst $\alpha \ge 0.025$ (outer cluster), but reducing $\alpha \le 0.01$ (five inner lines) resulted in a significant drop in the ROCAUCs.

\subsection{Multi-Source Queries}

In the multi-source query cases, both the inclusion and exclusion radii are well-defined. For each clip, we simply query the system using hyperellipsoids centered at the centroid of the target embedding and with radii $\tilde{\mathbf{r}} = (\mathbf{r} + \mathbf{r}^\perp)/2$. Unlike the single-source case, the inclusion radii for the multi-source case are always strictly non-zero, eliminating the need to iterate through multiple possible queries to obtain a good estimate.

\begin{figure}[t]
    \centering
    \includegraphics[width=\columnwidth]{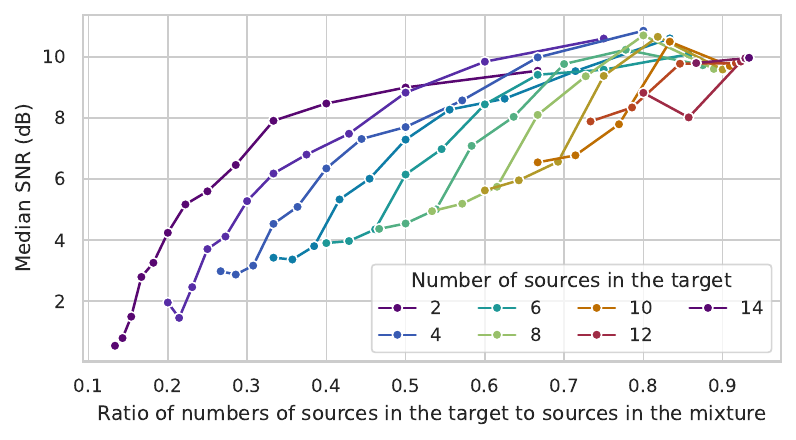}
    \caption{Multi-source queries: Median SNR (dB) by number of sources in the mixture and number of sources in the target.}
    \label{fig:ms-snr-abbrev}
\end{figure}

\begin{figure}[t]
    \centering
    \includegraphics[width=\columnwidth]{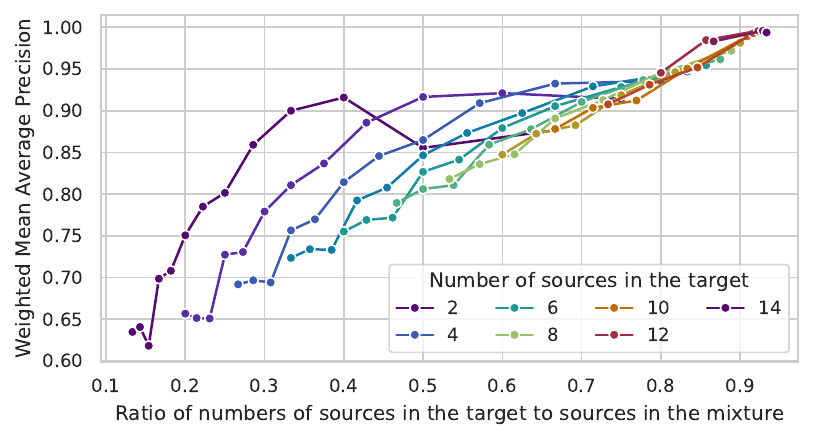}
    \caption{Multi-source queries: Weighted mean average precision by number of sources in the mixture and number of sources in the target.}
    \label{fig:ms-wmap}
\end{figure}

The median SNRs by the number of sources in the mixture and the number of sources in the target are shown in \Cref{fig:ms-snr-abbrev}. The full distributions are provided in \Cref{fig:ms-snr}. The corresponding weighted mean average precisions (mAP) by the number of sources in the mixture and the number of sources in the target are shown in \Cref{fig:ms-wmap}. The unweighted mAP plot is available in \Cref{fig:ms-map}, with a very similar trend as weighted mAP. Summary retrieval metrics are shown in \Cref{tab:ms-cls-abbrev}. The detailed retrieval metrics by source classes are provided in \Cref{tab:ms-cls}. 

\begin{table}[t]
    \centering
    \begin{tabularx}{\columnwidth}{X*{7}{S[table-format=1.2]}}
    \toprule 
    & {\textbf{AP}} & {\textbf{Acc.}} & {\textbf{Precision}} & {\textbf{Recall}} & {\textbf{F1}} \\
    \midrule
         \textbf{Macro Avg.} &   0.83 & 0.76 & 0.73 & 0.93 & 0.81 \\
\textbf{Micro Avg.} &  0.86 & 0.81 & 0.78 & 0.93 & 0.84 \\
\bottomrule
    \end{tabularx}
    \caption{Summary retrieval metrics for the proposed system in the multi-source query setup.}
    \label{tab:ms-cls-abbrev}
\end{table}

From \Cref{tab:ms-cls-abbrev}, the overall mAP of the system is \num{0.83} unweighted and \num{0.86} weighted. However, the system is clearly more performant in recall than precision. This, in source separation terms, indicates that most of the errors in the output are of an interference nature, rather than a distortion nature. 


By computing the ratio of the number of sources in the target to that of the mixture, we can see from \Cref{fig:ms-snr-abbrev} that, the larger the number of target sources relative to that of the mixture sources, the better the model performs in terms of SNR. At the same time, for a given number of target sources, the larger the number of sources in the mixture, the worse the model performs. This is further supported by \Cref{fig:ms-wmap} which shows a very similar trend in weighted mAP. This trend is somewhat unsurprising from a retrieval perspective, given that the number of relevant items relative to the total size of the collection affects the difficulty of a retrieval task. Given the generally strong performance of the system at a high ratio,

\section{Conclusion}
In this work, we propose a query-based music source separation system using hyperellipsoidal queries. This enables flexible control over both the ``timbre'' and the ``broadness'' of the query. The proposed method performs on par or better than state-of-the-art MSS systems, but is~---~unlike most state-of-the-art systems~---~not limited to a predefined set of target stems; instead, any source or a set of sources can be extracted as long as their embeddings can collectively be captured using a hyperellipsoid in the query embedding space. In future work, we hope to extend our work to support the use of negative queries, extraction of sources from disjoint regions, and automatic formation of region queries from examples, in order to better support real-world use cases of MSS in creative audio practices. 

\clearpage
\ifcamready

\fi

\small
\bibliographystyle{named}
\bibliography{references}

\clearpage


\appendix

\section*{SUPPLEMENTARY MATERIALS}

\setcounter{table}{0}
\renewcommand{\thetable}{\Alph{section}.\Roman{table}}

\setcounter{figure}{0}
\renewcommand{\thefigure}{\Alph{section}.\Roman{figure}}

\section{Additional Details on Data and Experimental Setup}

\subsection{Stem Abbreviation}

See \Cref{tab:instrument_abbreviations} for the stem abbreviations used in this work.

\begin{table}[t]
\centering
\begin{tabularx}{\columnwidth}{lX} 
    \toprule
    \textbf{Abbreviation} & \textbf{Stem} \\
    \midrule
    \textbf{Vx} & \textbf{Vocals}\\
    F Vx & Lead Female Singer \\ 
    M Vx & Lead Male Singer \\ 
    Bg Vx & Background Vocals \\ 
    Choir & Human Choir \\ 
    O Vx & Other Vocals \\ 
    \midrule
    \textbf{Bs} & \textbf{Bass}\\
    BGtr & Bass Guitar \\ 
    BSyn & Bass Synthesizer \\ 
    CBs & Contrabass / Double Bass \\ 
    Tba & Tuba \\ 
    Bsn & Bassoon \\ 
    \midrule
    \textbf{Dr} & \textbf{Drums} \\ 
    Snare & Snare Drum \\ 
    Kick & Kick Drum \\ 
    Toms & Toms \\ 
    Cymb & Cymbals \\ 
    OH & Overheads \\ 
    FADK & Full Acoustic Drumkit \\ 
    Dr Mach & Drum Machine \\ 
    \midrule
    \textbf{Perc} & \textbf{Percussion}\\
    A Perc & Atonal Percussion \\ 
    P Perc & Pitched Percussion \\ 
    \midrule
    \textbf{Gtr} & \textbf{Guitar}\\
    Dist EGtr & Distorted Electric Guitar \\ 
    Cln EGtr & Clean Electric Guitar \\ 
    LS/S Gtr & Lap Steel / Slide Guitar \\ 
    AGtr & Acoustic Guitar \\ 
    \midrule
    \textbf{Plucked} & \textbf{Plucked} \\
    O Plucked & Other Plucked \\ 
    \midrule
    \textbf{Pf} & \textbf{Piano}\\
    Grand Pf & Grand Piano \\ 
    EPf & Electric Piano \\ 
    \midrule
    \textbf{Keys} & \textbf{Keys} \\
    Org & Organ / Electric Organ \\ 
    Lead Syn & Synth Lead \\ 
    Pad Syn & Synth Pad \\ 
    O Keys & Other Keys \\ 
    \midrule
    \textbf{Winds} & \textbf{Winds}\\
    Brass & Brass \\ 
    Reeds & Reeds \\ 
    Fl & Flutes \\ 
    O Winds & Other Wind \\ 
    \midrule
    \textbf{Str} & \textbf{Strings}\\
    Vn & Violin \\ 
    Vns & Violin Section \\ 
    Va & Viola \\ 
    Vas & Viola Section \\ 
    Vc & Cello \\ 
    Vcs & Cello Section \\ 
    Strs & String Section \\ 
    O Str & Other Strings \\ 
    \midrule
    \textbf{Oth} & \textbf{Others}\\
    Fx & Fx \\ 
    Click & Click Track \\ 
    \bottomrule
\end{tabularx}
\caption{Stem Abbreviations}
\label{tab:instrument_abbreviations}
\end{table}

\subsection{Training Parametrization}
The model presented in this work is trained using an AdamW optimizer \citep{Loshchilov2019DecoupledWeightDecay} with an initial learning rate of \num{e-3} and a decay factor of \num{0.98} per epoch. Each epoch consists of 8192 training batches of 4, using a chunk size of \SI{10}{\second}. Each model is trained for 150 epochs on either an NVIDIA GeForce RTX 3090 or 4090 GPU (both \SI{24}{GB}). 

\clearpage

\section{Additional Results}

\subsection{Single-Source Queries}
See \Cref{tab:snr} for the median SNRs (dB) and RMS errors (dB) comparison between the single-source performance of the proposed method and Banquet (\textsc{q:all}, TE+DA variant). Note again that our method was evaluated clip-wise while Banquet was evaluated over the full track.

\begin{figure*}
    \centering
    \includegraphics[width=\linewidth]{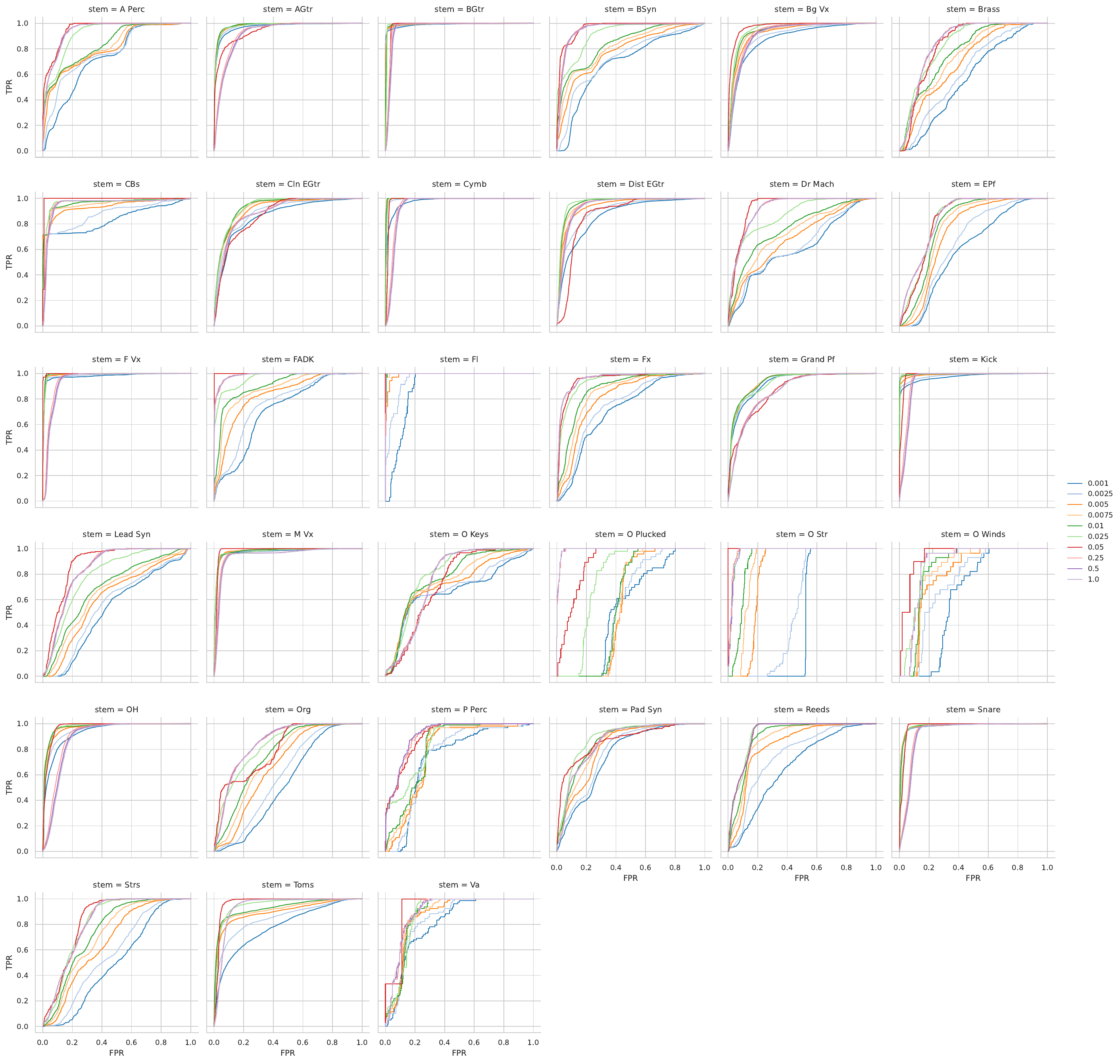}
    \caption{Single-source query ROC for each source class (``stem''), over different query scale factors $\alpha$.}
    \label{fig:all-roc}
\end{figure*}

See \Cref{fig:all-roc} for the ROC for each stem over different $\alpha$.

\begin{table}[t]
    \centering
    \begin{tabularx}{\columnwidth}{X*{4}{S[table-format=-3.2]}}
    \toprule
    & \multicolumn{2}{c}{\textbf{SNR (dB)}}
    & \multicolumn{2}{c}{\textbf{RMS Error (dB)}} \\
    \cmidrule(lr){2-3}\cmidrule(lr){4-5}
    &  {\textbf{Ours}} & {\textbf{Banquet}} 
    &  {\textbf{Ours}} & {\textbf{Banquet}}\\
    \textbf{``Stem''} 
    &  {(clip-wise)} & {(full)} 
    &  {(clip-wise)} & {(full)}\\
    \midrule
    F Vx & 9.8 & 9.9 & -0.4 & -0.2 \\
    M Vx & 8.2 & 7.8 & -0.4 & -0.4 \\
    Bg Vx & 0.4 & 0.1 & -0.4 & -4.9 \\
    \midrule
    BGtr & 10.5 & 10.8 & -0.3 & -0.2 \\
    BSyn & 0.9 & -0.1 & -8.0 & 0.2 \\
    CBs & 6.9 & {} & -0.6 & {} \\
    \midrule
    Snare & 6.0 & {} & -0.9 & {} \\
    Kick & 11.7 & {} & -0.3 & {} \\
    Toms & 4.9 & {} & -1.8 & {} \\
    Cymb & 4.2 & {} & -1.5 & {} \\
    OH & 2.6 & {} & -2.6 & {} \\
    FADK & 3.0 & {} & -2.6 & {} \\
    Dr Mach & 1.1 & {} & -3.4 & {} \\
    \midrule
    A Perc & -0.1 & {} & -4.8 & {} \\
    P Perc & 2.4 & -0.8 & -5.1 & -5.9 \\
    \midrule
    Dist EGtr & 4.0 & 2.7 & -2.1 & -1.1 \\
    Cln EGtr & 2.3 & 0.7 & -3.5 & -3.6 \\
    AGtr & 2.8 & 1.8 & -2.7 & -3.3 \\
    \midrule
    O Plucked & 0.0 & {} & -11.5 & {} \\
    \midrule
    Grand Pf & 2.8 & 2.3 & -2.8 & -1.7 \\
    EPf & 1.4 & 0.4 & -5.2 & -9.1 \\
    \midrule
    Org & 1.0 & 0.0 & -6.2 & -38.3 \\
    Lead Syn & 0.4 & 0.0 & -7.7 & -12.2 \\
    Pad Syn & 0.6 & 0.0 & -6.6 & -26.5 \\
    O Keys & 0.3 & {} & -6.8 & {} \\
    \midrule
    Brass & 1.6 & 0.0 & -5.7 & -28.9 \\
    Reeds & 3.2 & 0.0 & -3.7 & -19.4 \\
    Fl & 5.6 & {} & -2.0 & {} \\
    O Winds & 1.0 & {} & -6.4 & {} \\
    \midrule
    Va & 6.1 & {} & -1.6 & {} \\
    Strs & 0.9 & {} & -5.4 & {} \\
    O Str & 0.8 & {} & -4.4 & {} \\
    \midrule
    Fx & 0.5 & -0.2 & -7.3 & -14.3 \\
    \bottomrule
    \end{tabularx}
    \caption{Median SNRs (dB) and RMS errors (dB) comparison between the single-source performance of the proposed method and Banquet (\textsc{q:all}, TE+DA variant). Note again that our method was evaluated clip-wise while Banquet was evaluated over the full track.}
    \label{tab:snr}
\end{table}

\clearpage

\subsection{Multi-Source Queries}
See \Cref{fig:ms-snr} for the SNR (dB) distribution for multi-source queries by the number of sources in the mixture and the number of sources in the target.

\begin{figure}[h]
    \centering
    \includegraphics[width=\columnwidth]{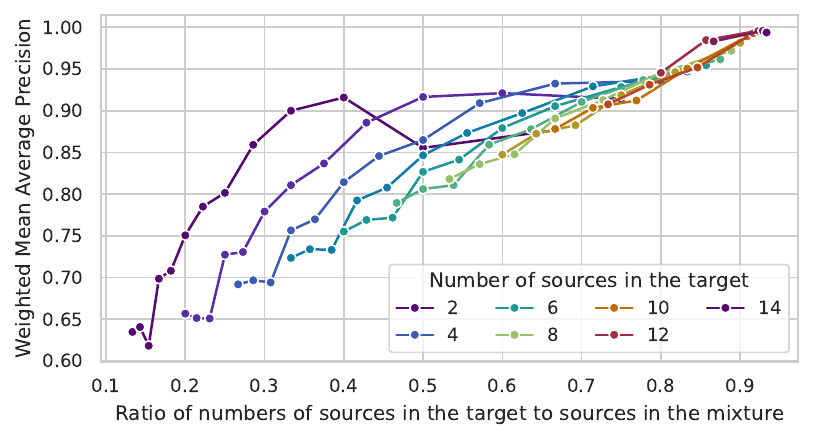}
    \caption{Weighted mean average precision by number of sources in the mixture and number of sources in the target.}
    \label{fig:ms-map}
\end{figure}

\begin{figure*}[t]
    \centering
    \includegraphics[width=\linewidth]{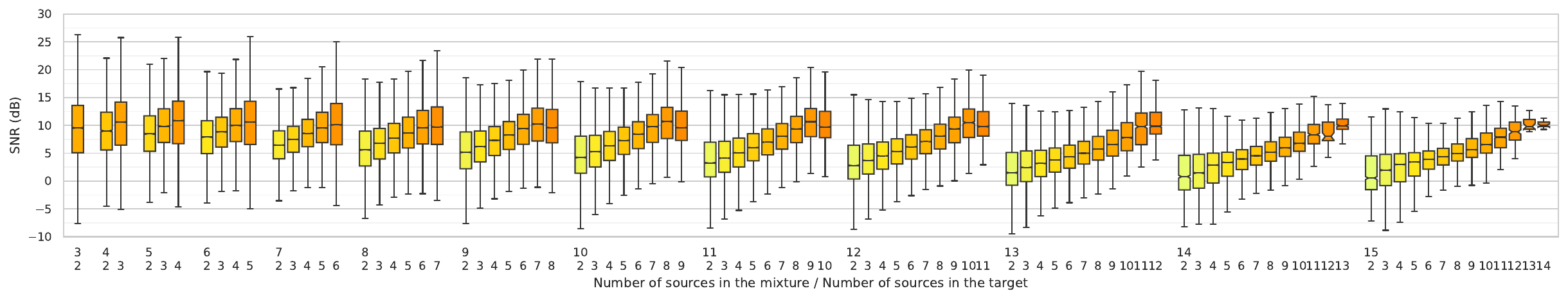}
    \caption{SNR distribution for multi-source queries by the number of sources in the mixture and the number of sources in the target.}
    \label{fig:ms-snr}
\end{figure*}

See \Cref{fig:ms-map} for the (unweighted) mean average precisions by the number of sources in the mixture and the number of sources in the target.

See \Cref{tab:ms-cls} for the detailed retrieval metrics over source classes. 
\newpage

\vfill

\begin{table*}[t]
    \centering
    \begin{tabular}{l*{8}{S[table-format=1.2]}}
    \toprule
    \textbf{Stem} & {\textbf{ROCAUC}} & {\textbf{PRAUC}} & {\textbf{AP}} & {\textbf{Accuracy}} & {\textbf{Precision}} & {\textbf{Recall}} & {\textbf{F1}} & {\textbf{Threshold}} \\
    \midrule
    F Vx & 0.93 & 0.92 & 0.92 & 0.88 & 0.85 & 0.95 & 0.90 & 0.22 \\
M Vx & 0.97 & 0.93 & 0.93 & 0.91 & 0.87 & 0.89 & 0.88 & 0.70 \\
Bg Vx & 0.90 & 0.87 & 0.87 & 0.82 & 0.77 & 0.90 & 0.83 & 0.55 \\
    \midrule
BGtr & 0.98 & 0.96 & 0.96 & 0.96 & 0.94 & 0.98 & 0.96 & 0.69 \\
BSyn & 0.81 & 0.84 & 0.84 & 0.74 & 0.69 & 0.96 & 0.80 & 0.18 \\
CBs & 0.97 & 0.97 & 0.97 & 0.91 & 0.89 & 0.96 & 0.93 & 0.43 \\
    \midrule
Snare & 0.96 & 0.94 & 0.94 & 0.91 & 0.88 & 0.94 & 0.91 & 0.61 \\
Kick & 0.98 & 0.97 & 0.97 & 0.94 & 0.93 & 0.96 & 0.94 & 0.58 \\
Toms & 0.92 & 0.91 & 0.91 & 0.86 & 0.83 & 0.92 & 0.87 & 0.58 \\
Cymb & 0.96 & 0.94 & 0.94 & 0.94 & 0.91 & 0.99 & 0.95 & 0.63 \\
OH & 0.91 & 0.87 & 0.87 & 0.86 & 0.82 & 0.94 & 0.88 & 0.51 \\
FADK & 0.73 & 0.69 & 0.69 & 0.68 & 0.62 & 0.91 & 0.74 & 0.17 \\
Dr Mach & 0.74 & 0.69 & 0.69 & 0.67 & 0.61 & 0.87 & 0.72 & 0.48 \\
    \midrule
A Perc & 0.84 & 0.84 & 0.84 & 0.75 & 0.70 & 0.91 & 0.79 & 0.49 \\
P Perc & 0.78 & 0.78 & 0.78 & 0.73 & 0.68 & 0.97 & 0.80 & 0.22 \\
    \midrule
Dist EGtr & 0.84 & 0.84 & 0.84 & 0.78 & 0.74 & 0.91 & 0.82 & 0.49 \\
Cln EGtr & 0.75 & 0.75 & 0.75 & 0.66 & 0.62 & 0.91 & 0.73 & 0.38 \\
AGtr & 0.87 & 0.86 & 0.86 & 0.78 & 0.72 & 0.85 & 0.78 & 0.58 \\
    \midrule
O Plucked & 0.67 & 0.74 & 0.74 & 0.62 & 0.61 & 0.93 & 0.74 & 0.23 \\
    \midrule
Grand Pf & 0.84 & 0.86 & 0.86 & 0.71 & 0.66 & 0.91 & 0.76 & 0.39 \\
EPf & 0.75 & 0.75 & 0.75 & 0.67 & 0.62 & 0.92 & 0.74 & 0.39 \\
    \midrule
Org & 0.69 & 0.72 & 0.72 & 0.61 & 0.59 & 0.94 & 0.72 & 0.27 \\
Lead Syn & 0.67 & 0.69 & 0.69 & 0.62 & 0.59 & 0.96 & 0.73 & 0.26 \\
Pad Syn & 0.87 & 0.88 & 0.88 & 0.80 & 0.79 & 0.86 & 0.82 & 0.61 \\
O Keys & 0.69 & 0.74 & 0.74 & 0.65 & 0.63 & 0.93 & 0.75 & 0.22 \\
    \midrule
Brass & 0.66 & 0.67 & 0.67 & 0.60 & 0.58 & 0.93 & 0.71 & 0.24 \\
Reeds & 0.85 & 0.87 & 0.87 & 0.75 & 0.71 & 0.93 & 0.81 & 0.14 \\
Fl & 0.85 & 0.81 & 0.81 & 0.78 & 0.72 & 0.93 & 0.81 & 0.28 \\
O Winds & 0.66 & 0.75 & 0.75 & 0.64 & 0.63 & 0.98 & 0.77 & 0.12 \\
    \midrule
Va & 0.85 & 0.88 & 0.88 & 0.75 & 0.73 & 0.88 & 0.79 & 0.63 \\
Strs & 0.68 & 0.62 & 0.62 & 0.61 & 0.57 & 0.95 & 0.71 & 0.38 \\
O Str & 0.93 & 0.96 & 0.96 & 0.84 & 0.85 & 0.91 & 0.88 & 0.50 \\
    \midrule
Fx & 0.76 & 0.79 & 0.79 & 0.67 & 0.63 & 0.90 & 0.74 & 0.40 \\
\midrule
\textbf{Macro Average} &  0.83 & 0.83 & 0.83 & 0.76 & 0.73 & 0.93 & 0.81 \\
\textbf{Micro Average} &  0.87 & 0.86 & 0.86 & 0.81 & 0.78 & 0.93 & 0.84 \\
\bottomrule
    \end{tabular}
    \caption{Retrieval metrics by source classes. }
    \label{tab:ms-cls}
\end{table*}

\null
\vfill

\newpage

\end{document}